\documentstyle[12pt]{article}
\begin{document}
\begin{flushright}
IITM-TH-94-02\\
\vspace{0.2cm}
May 1994 ~~~~~~~~
\end{flushright}

\vspace{1cm}

\baselineskip=24pt

\begin{center}
{{\Large \bf The \ geometry \ of \ electric \ charge: \\
The charge characteristic class}}\\
\vspace{0.3cm}
{by}\\
\vspace{0.3cm}
\bf {{G.H.Gadiyar}\\
{Department of Mathematics, Indian Institute of Technology,}\\
{Madras 600 036, INDIA.}}
\end{center}

\vspace{2cm}

\noindent {\bf Abstract.} It is well known that magnetic monopoles are
related to the  first Chern class. In this note electric charge is used
to construct an analogous characteristic class: the charge class.

\newpage

The intimate relationship between geometry and physics is now accepted by
all. A prime example of this relationship is the Yang-Mills theory. As is
well known the equations of this theory are
$$
F  =  D_A A \quad , \quad D_A F  =  0
$$
and
$$
D_A*F  =  *J \quad , \quad D_A*J  =  0
$$
where
$A$ is the connection,
$F$ is the curvature  (for mathematicians)
and
$A$ is the potential and
$F$ is the field strength  (for physicists).
$J$ is called the source by the physicists. Mathematicians
especially differential geometers set $J$ to be zero thus simplifying
the equations.

The question raised and answered in this article is what is the
geometric significance of $J$?  The answer to the question leads to the
construction of a new characteristic class: the charge class.

The key idea is to use the second set of Yang-Mills equations and
mimic the arguments usually carried out with the first set of
equations.

Recall the usual arguments for proving properties of characteristic
classes. These are outlined in [1]. One has to take a polynomial in $F$ defined
by
$$
Det ( tI~+~ i\frac{F}{2\pi}) ~=~ \sum_j t^j P_{m-j}(F) \ .
$$
Then $C_i(P) ~=~ P_i(F)$ are called Chern classes.

$F$ is a two-form. So $ P_i(F)$ is a $2i$ form. Then to show $P_i$ belongs to
$H^{2i} (M,R)$, two results have to be established.
\newline (i) \,$P_i(F)$ is closed
\newline and
\newline (ii) \,$P_i(F)$ is independent of the connection $A$ used to
compute $F$.

The proof that follows is modelled on the usual arguments. The main
difference is, to repeat, instead of using the first set set, the
second set of Yang-Mills equations is used.

The object considered in the case of the monopole is $Tr \, F$, the
first Chern class. Here by analogy the object considered in the case of
electric charge is $Tr \, *J$. This is called the charge class in this
paper. Notice
that $*J$ is a three-form. So one should get an element of $H^3(M,R)$.

Now conservation of current reads
$$
D_A*J~=~0 \ .
$$
Using the definition of $D_A*J ~=~d*J~+~A \wedge *J ~-~ *J \wedge A$
and \linebreak $Tr \, A\wedge B ~=~ (-1)^{pq}\, Tr \, B\wedge A $ if $A$ is a
$p$
form and $B$ is a $q$ form, it
\linebreak  follows that
$$
d \, Tr \, *J ~=~0 \ .
$$
Hence $Tr \, *J $ is closed.

The next result to be established is that $Tr \, *J$ is independent
of the connection $A$ chosen. Again the proof follows standard lines.

Let $Tr \, *J $ and $Tr \, *J'$ be the charge classes corresponding to
connections $A$ and $A'$. To prove independence of the charge class
it has to be established that
$$
Tr \, *J ~-~ Tr \, *J' ~=~ d \eta
$$
for some $\eta$, then it follows that $Tr \, *J$ and $Tr \, *J'$
belong to the same cohomology class in $H^3(M,R)$.

Consider two connections $A$ and $A'$ and define a family of connections
\begin{eqnarray*}
A^t & = & A~+~t(A'~-~A)\\
& = & A~+~t\, a
\end{eqnarray*}
where $a~=~ A'~-~ A$.

Corresponding to $A^t$ will be $F^t$ and $J^t$. Note
$$
Tr \, *J ~-~ Tr \, *J' ~=~ \int^1_0 \frac{d}{dt} \, Tr \, (*J^t)dt \ .
$$
It is shown that
$ \displaystyle{\frac{d}{dt}}\, Tr \, *J^t $ is exact for $0\, \leq \,t \, \leq
\, 1$,
that is,
$$
\frac {d}{dt}\, Tr \, *J^t~=~ d\,\theta (t) \ , \quad 0\, \leq \, t \, \leq \,
1 \ .
$$
Thus
\begin{eqnarray*}
Tr \, *J ~-~ Tr \, *J' & = & \int^1_0 d \, \theta (t) \, dt \\
& = & d \int^1_0 \theta (t) \, dt\\
& = & d \eta  \hspace{7cm} (1)
\end{eqnarray*}
where $ \displaystyle {\eta}~=~\displaystyle {\int^1_0 \theta (t) \, dt}$.

The proof that $\displaystyle {\frac{d}{dt}\, Tr \, *J^t}$ is exact for $0~\leq
{}~t~\leq ~1$ is displayed below. It is sufficient to show
$\displaystyle {{\frac{d}{dt}\, Tr \, *J^t} \vert _{t~=~0}}$ is exact.

For if $\displaystyle{\frac{d}{dt}} \, Tr \, *J(t) $ is exact for $ t ~ = ~ 0$,
it is also
exact for for any subsequent value of $t$, called $t'$. For the interval [0,1]
can be replaced by [t,1] and replace $A$ and $A'$ by $A^t$ and $A'$.
$$
\frac{d}{dt} \, Tr \, *J^t ~=~ Tr \, *\frac {d}{dt} \, J^t \ .
$$
But  $D_{A^t} \,*F^t~=~ *J^t$ and
\begin{eqnarray*}
F^t & = & d A^t ~+~ A^t \wedge A^t\\
& = & d(A~+~ta)~+~(A~+~ta) \wedge (A~+~ta)\\
& = & F~+~ t(da~+~A\wedge a~+~a \wedge A)~+~t^2 \, a\wedge a \ .
\end{eqnarray*}
So
\begin{eqnarray*}
\frac{d}{dt} \, Tr \, *J^t & = & \frac{d}{dt} \, Tr \, D_{A^t} \, *F^t\\
& = & \frac{d}{dt} \, Tr \, (d*F^t~+~A^t \wedge F^t~-~ F^t \wedge A^t)\\
& = & \frac{d}{dt} \, Tr \, d*F^t\\
& = & Tr \, d* \frac{dF^t}{dt}\\
& = & Tr \, d*(da~+~A\wedge a~+~a \wedge A)\\
& = & d \, Tr \, *(da~+~ A\wedge a~+~a \wedge A)\\
& = & d \, Tr \, *da \ .
\end{eqnarray*}
Thus $\displaystyle{\frac{d}{dt}} \, Tr \, *J^t ~=~ d\, Tr \, *da$.
Using (1) it follows that
$$
Tr \, *J ~-~ Tr \, *J' ~=~ d \eta \ .
$$
Thus we have established
\newline
\noindent (i) \quad $Tr \, *J$ is closed.
\newline
\noindent (ii)\quad $Tr \, *J$ is independent of connection $A$ used to
compute $F$.

The situation is clear in the case of electro magnetism. The equations then
collapse to
$$
\begin{array}{lcl}
F~=~d A \quad &,& \quad dF~=~0 \\
\\
*J~=~d*F \quad &,& \quad d*J~=~0 \ .
\end{array}
$$

The idea outlined here is as follows:
The second set of equations is the same as the first set, except that
$A$ is replaced by $*F$ and $F$ is replaced by $*J$.
Note however that these are forms of different degree.

A monopole is an appropriate configuration of $A$ (with a Dirac `string'
or singularity) and is related to $Tr \, F$ or the first Chern class.

Corresponding to $Tr \, *J$ or charge class is an appropriate configuration
of $*F$ (with some singularity). This corresponds to nontrivial
$H^3(M,R)$. In short we have used the current conservation law
instead of the Bianchi identity to prove the result. This gives some idea
as how to interpret the current $J$ in Yang-Mills theory.

\noindent {\bf {Reference}}.

\noindent [1]. C.Nash and S.Sen,  {\it {Topology and Geometry for physicists
}}, Academic Press, 1983.

\end{document}